\documentclass[12pt]{article}

\usepackage[centertags]{amsmath}
\usepackage{amsfonts}
\providecommand{\abs}[1]{\lvert #1 \rvert}
\providecommand{\absb}[1]{\Big\lvert #1 \Big\rvert}

\allowdisplaybreaks[1]
\author{Juan Racker$^{1,3}$\footnote{racker@ecm.ub.es - Present address: Departament d'Estructura i Constituents de la Mat\`eria and 
ICCUB, Institut de Ci\`encies del Cosmos,  Universitat
  de Barcelona, Diagonal 647, 08028 Barcelona, Spain.}, Pablo
  Sisterna$^2$\footnote{sisterna@mdp.edu.ar},$\>$
  and Hector Vucetich$^3$\footnote{vucetich@fcaglp.unlp.edu.ar} \\[5pt]
  $^1${\normalsize \it CONICET, Centro At\'omico Bariloche, Avenida
    Bustillo 9500 (8400) Bariloche,} \\ [-2pt] {\normalsize \it
    Argentina} \\[-2pt]
  $^2${\normalsize \it Facultad de Ciencias Exactas y Naturales,
    Universidad
    Nacional de Mar del Plata,} \\ [-2pt] {\normalsize \it Funes 3350
    (7600) Mar del Plata, Argentina } \\ 
  [-2pt] $^3${\normalsize \it Facultad de Ciencias Astron\'omicas y
    Geof\'{\i}sicas, Universidad Nacional de La Plata,} \\ [-2pt]
  {\normalsize \it Paseo del Bosque S/N (1900) La Plata, Argentina }}

\title{Thermodynamics in Variable Speed of Light Theories}
\begin{document}
\maketitle
\begin{abstract}
  The perfect fluid in the context of a covariant variable speed of
  light theory proposed by J. Magueijo is studied. On the one
  hand the modified first law of thermodynamics together with a recipe
  to obtain equations of state are obtained. On the other hand the
  Newtonian limit is performed to obtain the nonrelativistic
  hydrostatic equilibrium equation for the theory. The results
  obtained are used to determine the time variation of the radius of
  Mercury induced by the variability of the speed of light ($c$), and
  the scalar contribution to the luminosity of white dwarfs. Using a
  bound for the change of that radius and combining it with an
  upper limit for the variation of the fine structure constant, a
  bound on the time variation of $c$ is set. An independent bound is
  obtained from luminosity estimates for Stein 2015B.

\end{abstract}
\section{Introduction}

There are several very different motivations for studying the
variation of fundamental constants. The coincidence of large
dimensionless numbers arising from the combination of different
physical constants led Dirac to propose the large number hypothesis
and predict a time variation of them \cite{Dirac37},
\cite{Dirac38}. Theories with varying constants can also be a way of
implementing Mach's principle. Extra-dimensional theories like
superstring or Kaluza-Klein theories reduce in the low energy limit to
effective theories in which the fundamental constants may vary in
space and time.

Although no variation has been found in most experiments and
observations performed up to date, the results from analysis of spectra
from high redshift quasar absorption systems remain
controversial. Some works have reported a variation in the fine
structure constant \cite{Webb99}, \cite{Webb01}, \cite{Murphy01a},
\cite{Murphy01b}, \cite{Murphy03bis}, \cite{Lev07}, while other
studies give null results \cite{Mart03}, \cite{Quast04}, \cite{Bah04},
\cite{Sri04}. Besides the motivations mentioned above, variable speed
of light (VSL)
theories are interesting because they could solve several
cosmological puzzles \cite{Moffat93}, \cite{AlbMag99},
\cite{Barrow99}.

In this work we study the thermodynamics and Newtonian limit of the
varying speed of light theory developed by J. Magueijo
\cite{Mag00}. This theory was an improvement of a previous version
\cite{AlbMag99}, in order to account more carefully of local Lorentz
invariance and the dynamics of the scalar field behind $c$'s
variation. The theory here considered has been strongly criticized in
\cite{Ellis2003}, responded in \cite{Magueijo2003}, criticized again in
\cite{EllisUzan2003}, \cite{Ellis2007}, and answered back in
\cite{MagueijoMoffat2007}. As we point out at the end of the paper, in
Magueijo's theory the interaction between matter and the scalar field
is not explicitly shown in full detail. This can be seen in the
effective quantum creation of particles (see Sec.~VI of
Ref.~\cite{Mag00}), even in the absence of an explicit interaction term
in the matter Lagrangian ($b=0$ case; see below).  Nevertheless, being
a theory that gathered a lot of attention in the recent past, we
consider that it is very suitable as a first application of the
general framework developed in this work, which we plan to apply to
other (formally less controversial) theories such as bimetric
theories, in future publications.

After a summary of this VSL theory (Sec.~2) and of a
Lagrangian approach to describe perfect fluids (Sec.~3), the first
law of thermodynamics and a recipe for obtaining equations of state
are derived (Sec.~4). It is shown that the field associated with
the variation of the speed of light can formally be considered
as a new thermodynamic variable. Regarding this point we note that
this field has two properties that are not common in thermodynamic
variables: local universality and long scale variation. This has been
remarked in a work in the context of another theory with
variation of physical constants \cite{Canuto79}. In Sec.~5 we
perform the Newtonian limit and in Sec.~6 we apply all the results to
study the evolution of the radius of Mercury and derive a bound on the
time variation of $c$. Section 7 devotes to the estimation of the
scalar contribution to the luminosity of a white dwarf, obtaining a
stringent upper bound for the variation of $c$. In Sec.~8 we state
our conclusions and we leave for the Appendix some results concerning
the space and time dependence of the scalar field.

\section{Brief description of the VSL theory}

In the covariant and locally Lorentz invariant VSL theory proposed by
Magueijo $c$ plays three different roles:
\begin{itemize}
\item[{\bf a)}] \emph{$c$ is a dynamical field:}\\
  The spacetime variations of $c$ are represented by an adimensional
  scalar field $\psi$, so that
  \begin{equation}
    c=c_0 e^\psi,
  \end{equation}
  where $ c_0 $ is a constant.

  The general relativity (GR) action is modified and becomes
  \begin{equation}
    \label{eq:accionMag}
    I=\int d^4 x \; \sqrt{-g} \; \left(e^{a \psi}(R -2 \Lambda  - \kappa \nabla_\mu
      \psi \nabla^\mu \psi) + \frac{16 \pi G}{c_{0}^{\; 4}} e^{b \psi}
      \mathcal{L}_m \right) .
  \end{equation}
  The metric is taken to be $ \eta_{\mu \nu} = Diag (-1,1,1,1) $. $
  \mathcal{L}_m $ is the matter Lagrangian and $\kappa, a,$ and $b$ are
  three parameters of the theory. We will take $a-b=4$ as in
  \cite{Mag00}.  The matter Lagrangian is required to have no explicit
  dependence on $c$ (\emph{minimal coupling condition}), which leads to
  the second role played by $c$.

\item[{\bf b)}] \emph{$c$ parametrizes the variations of the other
    ``constants'':} The minimal coupling condition fixes the scaling
  with $c$ of all the Lagrangian parameters up to the $\hbar (c)$
  dependence, which is taken to be $ \hbar \propto c^{q-1} $, where $q$ is the fourth
  parameter of the theory. For example, since the Compton wavelength
  appears in the Lagrangian of a quantum particle, the dependence on
  $c$ of the mass will be $ m \propto \hbar / c \propto c^{q-2}$. In a similar fashion
  it can be determined that the charge of a quantum particle ($e$),
  the Bohr radius ($r_b$), and the fine structure constant ($\alpha$) are
  proportional to $c^q, c^{-q},$ and $ c^q$, respectively.

\item[{\bf c)}] \emph{$c$ is an integrating and conversion factor:}
  The theory is covariant and locally Lorentz invariant in a
  generalized way explained in \cite{Mag00}. The key point is the use
  of an $x^0$ coordinate, instead of time, in all geometrical
  formulas. The main difference with constant $c$ theories is that
  local measurements of space and time ($dx$ and $dt$) are not
  generally differentials of a coordinate system (the partial
  derivatives do not commute). Nevertheless it is always possible to
  find integrating factors such that $ dt \, \psi^\beta $ and $ dx \, \psi^{\beta
    -1} $ are perfect differentials, with $ \beta $ another parameter which,
  however, will not appear in the equations of our work, because the
  calculations will be done using the $x^0$ coordinate.
\end{itemize}

Varying the action $\eqref{eq:accionMag}$ with respect to the metric
and $\psi$ leads to the equations:
\begin{equation}
  \label{eq:Mg}
  \begin{split}
    G_{\mu \nu} + \Lambda g_{\mu\nu} &= \frac{8 \pi G}{c^4} T_{\mu\nu} + \kappa \left( \nabla_\mu \psi
      \nabla_\nu \psi - \frac{1}{2} g_{\mu\nu} \nabla_\delta \psi \nabla^\delta \psi \right) + \;\\ & \quad \;
    e^{-a \psi}\left(\nabla_\mu \nabla_\nu e^{a \psi} - g_{\mu\nu} \Box e^{a \psi} \right),
  \end{split}
\end{equation}
\begin{equation}
  \label{eq:Mpsi}
  \Box \psi + a \nabla_\mu \psi \nabla^\mu \psi = \frac{8 \pi G}{c^4 (2\kappa +
    3a^2)}\left(aT - 2a\rho_\Lambda c^2 - 2b \mathcal{L}_m \right) +
  \frac{1}{\kappa} \frac{d \bar{\Lambda}}{d \psi} \, ,
\end{equation} 
where $T^{\mu \nu}$ is the matter stress energy tensor and $T$ is its
trace. Besides, $\rho_\Lambda$ is the mass density corresponding to the cosmological
constant, $\rho_\Lambda = \frac{\Lambda c^2}{8 \pi G}$ and $\bar{\Lambda}$ is a linear
combination of the matter and geometric cosmological
constants, $\bar{\Lambda} = \Lambda + \frac{8 \pi G}{c^4} \Lambda_m$.

After applying the Bianchi identities to Eq.~$\eqref{eq:Mg}$ and using
Eq.~$\eqref{eq:Mpsi}$, an equation for the divergence of $ T^{\mu \nu} $ is
obtained:
\begin{equation}
  \label{eq:M3}
  T ^\nu{} _{\mu;\nu} = - \psi_{;\nu} T^\nu{}_\mu b +
  \psi_{;\mu} b \mathcal{L}_m - \Lambda _{m;\mu} \; .
\end{equation}
Note that matter energy is conserved only when $b=0$, in all other cases there is exchange of energy between matter and
  the $\psi$ field.

The presence of the $\psi$ field can modify the law of conservation of
the number of particles and the normalization condition for the
four-velocity, e.g. it is found that for a classical particle $U^2 =
U_0^{\, 2} \left( c/c_0\right) ^ {-b} \neq -1 $ \cite{Mag00}. Besides,
the energy density and the total energy of a body in hydrostatic
equilibrium will also vary if $c$ is not constant (we must distinguish
between energy density and total energy because the size of a body can
change in time as a result of the variation of $c$). In addition, the
$c$ dependence of the mass is different for a classical and a quantum
particle. Finally, the matter Lagrangian, which is not unique, appears
in the equations of the theory. To take into account all these effects
and ambiguities it is convenient to introduce four new parameters, $
q_1, q_2, q_3,$ and $ q_4 $:
\begin{eqnarray}
  \label{eq:normalizacion U}
  c^{q_1} U^{\mu} U_{\mu} = c_0^{q_1} U_0^2 =\mathcal{C}  &
  \begin{array}{c}
    \text { Generalized normalization }\\
    \text { of the four-velocity.}
  \end{array} \\
  \label{eq:conservacion n}
  (nc^{q_2}U^{\nu})_{;\nu} = 0  &
  \begin{array}{c}
    \text { Generalized conservation}\\
    \text { of particle number. }
  \end{array} \\
  \label{eq:variacion densidad}
  \rho = \rho_0 e^{q_3 \psi} &
  \begin{array}{c}
    \text { $c$ dependence of the energy density of}\\
    \text { a body in hydrostatic equilibrium.}
  \end{array}  \\
  \label{eq:variacion energia total}
  U = U_0 e^{q_4 \psi} &
  \begin{array}{c}
    \text { $c$ dependence of the total energy of}\\
    \text {a body in hydrostatic equilibrium.}
  \end{array}
\end{eqnarray}
The `` $0$ " subscript denotes the value of these quantities when
$ \psi = 0$.

\section{The perfect fluid Lagrangian}

The task of obtaining a Lagrangian for the perfect fluid is not a
trivial one due to the constraints imposed by the normalization of the
velocity and the conservation of the number of particles. A. H. Taub
gave one in 1954 \cite{Taub54} and another one was given by
B. F. Schutz in 1970 \cite{Schutz70}, using a different but equivalent
approach. Note that equivalent Lagrangians in usual theories
(i.e. differing in a divergence) may not be equivalent in Magueijo's
one, because of their explicit appearance in the evolution
equations. In our work we have used Schutz's Lagrangian, hence we present in this section a brief summary of his
approach and then we show how it can be used in the VSL theory under study.

\subsection{Schutz's Lagrangian}

Schutz uses a formulation of relativistic hydrodynamics based on the
utilization of 6 potentials to represent the velocity:
\begin{equation}
  U_{\nu} = \mu^{-1} \left( \phi_{,\nu} + \xi \beta_{,\nu} + \theta s_{,\nu} \right),
\end{equation}
where $\mu$ and $s$ are the specific enthalpy (enthalpy per unit mass)
and the specific entropy, respectively. The physical meaning of the
other potentials is also explored by Schutz.

The perfect fluid action is
\begin{equation}
  I=\int \left(R + \frac{16 \pi Gp}{c_{0}^{\; 4}} \right)(-g)^{1/2} \; d^4
  x \, ,\notag 
\end{equation}
where $p$ is the pressure of the fluid. Then the perfect fluid
Lagrangian is $\mathcal{L}_m = p$. The following steps must be
followed to vary the action:
\begin{enumerate}
\item Choose an equation of state for the fluid and write it in terms
  of $\mu$ and $s$: $p=p(\mu,s)$.
\item When varying the action make use of the thermodynamic relation
  $dp = \rho_m d\mu - \rho_m T ds$ (where $\rho_m$ is the rest mass density).
\item Define the four-velocity vector field $U_{\nu}$ in terms of 6
  scalar potentials: $ U_{\nu} = \mu^{-1} \left( \phi_{,\nu} + \xi \beta_{,\nu} + \theta
    s_{,\nu} \right)$.
\item The normalization of $U$ is taken into account before starting
  the variations. This is done expressing $\mu$ in terms of $ \phi , \xi , \beta
  , \theta , s,$ and $g^{\mu \nu} $: $ \mu^{2} = - g^{\mu \nu} \left( \phi_{,\mu} + \xi \beta_{,\mu} +
    \theta s_{,\mu} \right)$ $\left( \phi_{,\nu} + \xi \beta_{,\nu} + \theta s_{,\nu}
  \right)$. So the independent variables are $ \phi , \xi , \beta , \theta , s,$ and
  $ g^{\mu \nu} $. Any quantity appearing in the action must be considered
  a function of these variables.
\end{enumerate}
The Euler-Lagrange equations become
\begin{gather*}
  G_{\mu \nu} - \frac{8 \pi G}{c_{0}^{\; 4}} \left[ (\rho+p) U_\mu U_\nu + p
    g_{\mu\nu} \right] = 0 \, ,\\
  (\rho_m U^\nu)_{; \nu} = 0 \, ,\\
  U^\nu s_{,\nu} = 0, \quad U^\nu \theta_{,\nu} = T, \quad U^\nu \beta_{,\nu} = 0, \quad
  \text{and} \quad U^\nu \xi_{,\nu} = 0 \, .
\end{gather*}
The stress energy tensor is
\begin{equation}
  T^{\mu \nu} = \frac{2}{\sqrt{-g}} \frac{\delta\left(\sqrt{-g} \mathcal{L}_m
    \right)}{\delta g_{\mu \nu}} =  (\rho+p) U^\mu U^\nu + p g^{\mu\nu} \, .
\end{equation}
It is important to note that $ (\rho + p), U_\mu , p, \rho_m$, and $T$ were
defined in terms of $ \phi , \xi , \beta , \theta , s,$ and $ g^{\mu \nu} $ after
performing the variations, using the equations
\begin{gather*}
  p=p(\mu,s)\qquad \text{(equation of state)}, \\
  \rho_m = \left(\frac{\partial p}{\partial \mu} \right)_s, \quad T = \frac{-1}{\rho_m}
  \left(\frac{\partial p}{\partial s} \right)_\mu ,\\
  (\rho + p) = \rho_m \mu, \quad U_{\nu} = \mu^{-1} \left( \phi_{,\nu} + \xi \beta_{,\nu} + \theta
    s_{,\nu} \right).
\end{gather*}

\subsection{Use of Schutz's Lagrangian in Magueijo's theory}

Using Schutz's Lagrangian in the action $\eqref{eq:accionMag}$ and
varying it with respect to $ \phi , \xi , \beta , \theta $, and $ s $ we obtain 
\begin{gather}
  \label{eq:conservacion masa}
  (e^{b\psi}\rho_m U^\nu)_{; \nu} = 0, \\
  U^\nu s_{,\nu} = 0, \\
  U^\nu \theta_{,\nu} = T, \\
  U^\nu \beta_{,\nu} = 0, \\
  U^\nu \xi_{,\nu} = 0.
\end{gather}
Moreover, varying $\mathcal{L}_m$ with respect to $g^{\mu \nu}$ leads to
\begin{equation}
  \label{eq:tensorEM}
  T^{\mu\nu} = (\rho+p) U^\mu U^\nu + p g^{\mu\nu} \, ,
\end{equation}
where $ \rho , p, \rho_m,$ and $U^\mu $ are defined from $ \phi , \xi , \beta , \theta $, and $ s $
exactly in the same way as in Schutz's theory. In the VSL theory they
can depend on $\psi$, but they will coincide with the usual quantities when $\psi = 0$. Besides, with these definitions of $ \rho, p,$ and $U^\mu $,
the stress energy tensor in the VSL theory has the same form as the
usual one for a perfect fluid with energy density $\rho$ and pressure $p$
(both quantities being measured in a momentarily comoving reference
frame). These facts make it reasonable to consider $ \rho , p, \rho_m $, and
$ U^\mu $ as the energy density, pressure, rest mass density, and four-velocity of the perfect fluid in the VSL theory that is being studied.

Note that by definition $U^\mu U_\mu = -1$, so $\mathbf{q_1 = 0}$. This
is different from Magueijo's result for a classical particle, $ e^{b\psi}
U^\mu U_\mu = $ constant, where the definition of the velocity is $ U^\mu =
\frac{dx^\mu}{d\lambda} = \frac{dx^\mu}{cd\tau}$. Although there is no contradiction
with this, one has to be careful with the interpretations given to
$U^\mu$.


\section{Thermodynamics}
\subsection{First law}

When energy is conserved, the divergence of $T^{\mu \nu}$
[Eq.~\eqref{eq:M3}] is zero and the first law of thermodynamics is
obtained projecting along $U^{\mu} $.  We will do the same here but
using $ c^{q_1} U^\mu $ as the projector. Although using Schutz's
Lagrangian leads to work with a four-velocity normalized to $ -1 \;
(q_1 = 0) $, the calculations in this section will be done with an
arbitrary $ q_1 $. The motivation is to obtain equations valid even
when that condition is not satisfied.

The results of this section will be applied to systems whose scales
are much smaller than cosmological scales, so we take $\Lambda_m=0$
\footnote{On the other hand, the cosmological constant is important
  for the evolution of $\psi$.}.  Projecting $ T^{\nu}_{\, \mu ; \nu} $ along $
c^{q_1} U^\mu $ and using Eqs.~$\eqref{eq:M3}$, $\eqref{eq:tensorEM}$,
$\eqref{eq:normalizacion U}$, and $\eqref{eq:conservacion n}$ together
with Schutz's Lagrangian leads to \footnote{If $\Lambda_m\neq 0$ but depends
  only on $\psi$, the equation will be valid after adding the term $-
  \frac{c^{q_1}}{\mathcal{C}} \frac{d\Lambda_m}{d\psi} d\psi$ to the right-hand
  side.}
\begin{equation}
  \label{eq:ley1b}
  d\rho + \left[1 + \frac{c^{q_1}}{\mathcal{C}}\right]dp - \frac{(\rho+p)}{n}dn =
  (\rho+p)\left(\frac{q_1}{2} + q_2 - b \right)d\psi \, .
\end{equation}
It is convenient to express this relation in terms of the specific
thermodynamic variables, $\displaystyle v=\frac{V}{N}=\frac{1}{n}$,
$\displaystyle u=\rho v=\frac{U}{N} $, and $ \displaystyle s= \frac{S}{N} $
($V$, $U$, and $S$ are the total volume, total energy, and total entropy
of a system containing $N$ particles):
\begin{equation}
  \label{eq:ley1c}
  du + p dv -v(\rho+p)\left(\frac{q_1}{2} + q_2 - b \right)d\psi + v\left[1 +
    \frac{c^{q_1}}{\mathcal{C}}\right]dp = 0 \, .
\end{equation}
The first two terms are the only ones appearing in GR. The third term
shows that $\psi$ formally plays the role of a new thermodynamic variable
(changes in $\psi$ cause changes in the internal energy). Finally, the fourth term
involves the pressure, which is not an independent variable (up to
this point the independent variables are taken to be $ v $ and $ \psi
$). The expression is valid only for thermodynamic processes which do not
involve heat transfer, while in a more general situation, with an amount $d Q$ of heat being transferred, the equation becomes
\begin{equation}
  \label{eq:ley1termo}
  du + p dv -v(\rho+p)\left(\frac{q_1}{2} + q_2 - b \right)d\psi + v\left[1 +
    \frac{c^{q_1}}{\mathcal{C}}\right]dp = d Q \, .
\end{equation}
This is the modified first law of thermodynamics in the VSL theory.

Incorporating Caratheodory's principle it can be shown that there
exists an integrating factor for $d Q$ in Eq.~\eqref{eq:ley1termo}
\cite{Chandra57}. Defining $ \frac{1}{T} $ as the integrating factor
and $d s= \frac{d Q}{T}$, Eq.~\eqref{eq:ley1termo} becomes
\begin{equation}
  \label{eq:ley1d}
  du + v\left[1 + \frac{c^{q_1}}{\mathcal{C}}\right]dp +p dv -
  v(\rho+p)\left(\frac{q_1}{2} + q_2 - b \right)d\psi = Tds \, ,
\end{equation}
where $s$ (identified with the specific entropy of the system) and $T$ are two thermodynamic variables. 
To go on, it is convenient to introduce the function $f(\psi) = 1 +
\frac{c^{q_1}}{\mathcal{C}}= 1 +
\frac{c_0^{q_1}}{\mathcal{C}}e^{q_1\psi}$. We take as the time of
reference the present epoch, so that $U_0^2=-1$ (which is the usual normalization of the
four-velocity) and $c_0^{q_1}=c_{today}^{q_1}$. Then from Eq.~\eqref{eq:normalizacion U} it follows that
$\mathcal{C}=-c_0^{q_1}$ and hence $f(\psi)=1 - e^{q_1\psi}$.
After choosing $s$, $v$, and $\psi$ as the independent variables and
writing $p$ in terms of them, an expression for the first law of thermodynamics
involving only state variables is finally obtained:
\begin{equation}
  \label{eq:ley1M}
  \begin{split} \\
    \quad du &=-\left(p+vf(\psi)\frac{\partial p}{\partial v}\right)dv +
    \left(T-vf(\psi)\frac{\partial p}{\partial s}\right)ds \: +\quad\\ \\
    &\qquad\left((u+pv)\left(\frac{q_1}{2} + q_2 - b
      \right)-vf(\psi)\frac{\partial p}{\partial \psi}\right)d \psi \; .
  \end{split}
\end{equation}

\subsection{Equations of state}

The first partial derivatives of the specific internal energy can be
obtained directly from Eq.~$\eqref{eq:ley1M}$. There are three different
equalities between the mixed partial derivatives which impose some
restrictions on the functional dependence of $p$ and $T$ on $\psi$:
\begin{equation}
  \label{eq:derivadas sv}
  \left( f(\psi)-1 \right) \frac{\partial p}{\partial s} =
  \frac{\partial T}{\partial v},
\end{equation}
\begin{equation}
  \label{eq:derivadas vpsi}
  \frac{\partial p}{\partial \psi} = b_1 v \frac{\partial p}{\partial v},
\end{equation}
\begin{equation}
  \label{eq:derivadas spsi}
  \frac{\partial T}{\partial \psi} = b_2 T + b_1 v \frac{\partial T}{\partial v}
  \, ,
\end{equation}
with $b_1 = \frac{q_1}{2} - q_2 + b$ and $b_2 = \frac{q_1}{2} + q_2 -
b$.

Several observations corresponding to different epochs in the history
of the Universe show that the $\alpha$ variation has been very small (or
zero) \cite{Uzan03}, so the field $\psi$ must also be very small. Then,
it makes sense to express the pressure as a power expansion in $\psi$~\footnote{The $\psi$ dependence of the temperature can be treated in a
  similar fashion.}:
\begin{equation}
  p = \sum_{k=0}^{\infty} p_k (v,s) \psi^k \, .
\end{equation}
After replacing this series in Eq.~\eqref{eq:derivadas vpsi} and equaling
terms with the same power of $\psi$ a recurrent formula for the
coefficients $p_k$ is obtained:
\begin{equation}
  \displaystyle p_{k+1}(v,s) = \frac{b_1v}{k+1} \frac{\partial p_k
    (v,s)}{\partial v} \, ,
\end{equation}
where $p_0(v,s)$ is the pressure as a function of $v$ and $s$ for
$\psi=0$ and therefore it is obtained from the usual theories in which $c$ is
constant.

Working to first order in $\psi$ we arrive at the following expression
for the functional dependence of $p$ on $v, s$, and $\psi$:
\begin{equation}
  \label{eq:presion}
  p \simeq  p_0(v,s) + b_1v \frac{\partial p_0}{\partial v} \psi \, .
\end{equation}

\section{Newtonian limit}

The Newtonian limit of this VSL theory can be obtained following the
same steps as those used in GR. Although in the Newtonian limit time derivatives are negligible
with respect to the spatial ones, special considerations are needed for the derivatives of $\psi$. For example, if the spatial extension of a system is
small compared to the scales associated with the spatial variations of $\psi$ (which are cosmological scales) and one is interested in following its time
evolution, time derivatives will be more interesting than the spatial
ones. This will be the case in the following sections, where we will
determine the evolution of the planetary radii and the luminosity of
white dwarfs. We also remind that, as explained before, the
cosmological constant is not included in our calculations. Finally, since the $\psi$ field must be very small, we will work to first order in $\psi$.

The condition of weak gravitational field allows to choose nearly
Lorentz coordinates in which
\begin{equation}
  \label{eq:aprox g}
  g_{\mu \nu} = \eta_{\mu \nu} + h_{\mu \nu}, \qquad \text{with} \; \abs{h_{\mu \nu}} \ll 1 \, .
\end{equation}
Using Eqs.~$\eqref{eq:Mg}$ and $\eqref{eq:Mpsi}$, the Lagrangian $\mathcal{L}_m = p$ and the fact that $ a \nabla_0
\nabla^0 \psi $ is negligible (see the Appendix), the following gravitational
potential equation is obtained:
\begin{equation}
  \label{eq:h00}
  - \nabla^2 h_{00} = \frac{8\pi \tilde{G}}{c^4}\; T_{00} \, ,
\end{equation}
where $\tilde{G} = G \left(1 + \frac{a^2}{2\kappa + 3a^2} \right)$ is an
effective gravitational constant. In the nonrelativistic limit of GR
$h_{00}$ is identified with $-2 \frac{\phi}{c^2}$ to arrive at Newton's
gravitational potential equation, instead here we will go on working with
$h_{00}$ due to the presence of the $c^2$ factor in that
identification.

The Euler's equations for this theory are obtained projecting
Eq.~$\eqref{eq:M3}$ perpendicularly to $U^\mu$ with the projector $ g^{\mu
  \alpha} + U^\mu U^\alpha$ (here we have taken $ q_1 = 0 $ from the beginning):
\begin{equation}
  - \frac{1}{2} T_{00} h_{00,i} + p_{,i} + \rho U^\nu U_{i,\nu} + \rho U_0 U^j
  g_{0j,i} = 0 \, .
\end{equation}
For quasistatical situations ($ U_i \approx 0 $) the equation becomes
\begin{equation}
  \label{eq:gradiente p estatico}
  \frac{1}{2} T_{00} \nabla h_{00} = \nabla p \, .
\end{equation}

The hydrostatic equilibrium equation follows after combining
Eqs.~$\eqref{eq:h00}$ and $\eqref{eq:gradiente p estatico}$. For a system
with spherical symmetry the result is
\begin{equation}
  \label{eq:equilibrio hidrostatico 2}
  \frac{dp}{dr} = - \frac{\tilde{G}}{c^4} \frac{\rho (r) U(r)}{r^2} \, ,
\end{equation}
where $U(r)$ is the total energy inside the sphere of radius $ r
$ and $\frac{d}{dr}$ must be understood as a spatial derivative at
constant time. To obtain this equation it is necessary to consider $c$ as
a constant in the integral $\int_0^r \frac{4 \pi r^2 \rho}{c^4} dr$. This can
be done because the spatial variations of $\psi$ are negligible in
noncosmological scales, as is demonstrated in the Appendix.

\section{Evolution of the radius of Mercury and a bound for the
  variation of $c$}

The radius of a planet is determined with the hydrostatic equilibrium
equation together with an equation of state and boundary
conditions. The presence of $\psi$ in these equations causes in general
time variations of the radius. On the other hand, the actual change in size of several
bodies of the Solar System have been estimated using different
topographical observations. For Mercury there is a stringent bound: its
radius has not changed more than 1 km in the last $ 3.9 \times 10^9 $
years \cite{Mc78}. This fact will allow us to obtain a bound for the
temporal variation of $\psi$.

In Sec.~6.1 we will show that the hydrostatic equilibrium equation is
equivalent to another equation in which the temporal dependence
resides only in the gravitational constant. Then it will be possible
to use the results of the work of McElhinny et al. and give a bound
for the quantity $ f(q_i) \dot{\psi} $, where $ f(q_i) $ is a function of
the parameters $ b_1, q_3 $, and $ q_4 $ (the dot denotes derivative
with respect to time). In Sec.~6.2 those parameters will be expressed in
terms of the parameters originally defined in \cite{Mag00}.  Finally,
in Sec.~6.3 the bound obtained for $ f(q_i) \dot{\psi} $ will be combined with
a bound for the variation of the fine structure constant to obtain an
upper limit for $|\dot{\psi}|$.

\subsection{Transformation into a variable $ G $ theory}

To solve the hydrostatic equilibrium equation it is necessary to have
an equation of state relating the pressure and density. It can be
obtained replacing the corresponding equation of state for the
constant $c$ case in Eq.~$\eqref{eq:presion}$. For Mercury it is
sufficient to work with a linear equation in $\rho$ \cite{Mc78}:
\begin{equation}
  \label{eq:estado}
  p_0 = K_{sur} \left( \frac{\rho_0}{\rho_{0 \, sur}}  - 1 \right) .
\end{equation}
The quantities with a ``0" subscript correspond to $\psi = 0$, $sur$ indicates evaluation in the surface of Mercury, and  $K_{sur}$ is
the superficial compressibility\footnote{In \cite{Mc78} a quotient
  between mass densities is used instead of $\frac{\rho_0}{\rho_{0 \,
      sur}}$, nevertheless in the Newtonian limit they are equal since $ \rho_{energy}
  \approx \rho_{mass} c^2 $.}.
Using Eqs.~$\eqref{eq:presion}$, $ \eqref{eq:estado} $, and the nonrelativistic expression $ \rho_0 \approx \frac{m_0 c_0^{\; 2}}{v}$ (with $ m_0 $ the average mass of a particle), the equation of state to be used is
obtained:
\begin{equation}
  \label{eq:presion Mer}
  p=K_{sur} \left( \frac{\rho_0}{\rho_{0 \, sur} } -1\right) - b_1 K_{sur} \psi
  \: \frac{\rho_0}{\rho_{0 \, sur}} \, .
\end{equation}

Replacing the equation of state~\eqref{eq:presion Mer} in Eq.~\eqref{eq:equilibrio hidrostatico 2}, using
the expression for $ \frac{\partial \psi}{\partial r} $ given in the Appendix and taking into account the
definitions of $ q_3 $ and $ q_4$ we get
\begin{equation}
  \begin{split}
    \frac{d p_0}{dr} = - \tilde{G} \: e^{\left(b_1 + q_3 + q_4 - 4
      \right) \psi} \: \frac{\rho_{m0} (r) M_0(r)}{r^2} - 2 b_1 \:
    \frac{K_{sur}}{\rho_{0 \, sur}} \: \frac{a}{2\kappa + 3a^2} \: e^{b_1 \psi} G
    \: \frac{\rho_{m0} (r) M_0(r)}{r^2} \, ,
  \end{split}
\end{equation}
where $ \rho_{m0}(r) $ is the mass density for $ \psi = 0 $ and $ M_0(r) $ is the
mass contained within radius $ r $ (also for $ \psi = 0 $). The second
term of the right-hand side is small compared to the first
one\footnote{Assuming $b_1,a = O(1)$, the quotient between these terms
  is no larger than approximately $\frac{K_{sur}}{\rho_{0 \, sur}}$. Since $K_{sur} \approx 10^{12} \: \frac{\text{dynes}}{\text{cm}^2} $ and $ \rho_{0
    \, sur} = \rho_{m0 \, sur} \: c_0^{\; 2} \approx 4 \times 10^{21} \:
  \frac{\text{dynes}}{\text{cm}^2} $ \cite{SS74}, we see that $\frac{K_{sur}}{\rho_{0 \, sur}} \approx 10^{-9} .$ }%
, so it can be multiplied
by $ \: e^{\left ( q_2 + q_4 -4 \right) \psi}$ [which is equal to $1 +
O(\psi)$] and hence the equation becomes
\begin{equation}
  \label{eq:Mercurio1}
  \frac{d p_0}{dr} = - \bar{G} \: e^{\left(b_1 + q_3 + q_4 - 4 \right)
    \psi} \: 
  \frac{\rho_{m0} (r) M_0(r)}{r^2} \, ,
\end{equation}
with $\bar{G}$ the final effective gravitational constant,
\begin{equation}
  \label{eq:G raya}
  \bar{G} = G \left( 1 + \: \frac{a^2}{2\kappa + 3 a^2} + 2 b_1 \: 
    \frac{K_{sur}}{\rho_{0 \, sur}} \: \frac{a}{2\kappa + 3 a^2} \right) \, .
\end{equation}
The last term within the parenthesis depends on the particular
properties of the body and could have, in principle, observational
consequences. Nevertheless, its value is too small relative to the
second (and body independent) term for its effect to be measurable.

The method used to set a limit on the actual time variation of the
radius of Mercury is based on the observation of surface features
\cite{Mc78}.  A homologous change $ ( R \propto M^{1/3})$ cannot be detected
through this procedure since it scales all linear dimensions
equally. Neither can be observed a variation $ R \propto r_b $ ( $ r_b $ is
Bohr radius ), since a change in $ r_b $ implies a change of all
macroscopic dimensions in the same proportion. For these reasons it is
necessary to make a change of variables:
\begin{equation}
\label{eq:scaling}
  r^\ast = \frac{r}{r_b M^{1/3}} \, , \qquad M^\ast = \frac{M(r)}{M} =
  \frac{M_0(r)}{M_0} \, ,
\end{equation}
where $ M $ is the total mass of Mercury.
Taking also into account that $ m \propto c^{q-2}$ and $ r_b \propto c^{-q} $,
Eq.~\eqref{eq:Mercurio1} becomes
\begin{gather}
  \label{eq:Mercurio2}
  \frac{d p_0}{d r^\ast} = - \bar{G}(t) \frac{M_0^{\; 2/3}}{r_{b \, 0}}
  \:
  \frac{\rho_{m0} (r^\ast) M^\ast_0(r^\ast)}{r^{\ast \, 2}} \, ,\\
  \intertext{with}
  \label{eq:G de t}
  \bar{G}(t) = \bar{G} \: e^{\left(b_1 + q_3 + q_4 - 4 + 2/3 q + 2/3
    \right) \psi(t)} 
\end{gather}
and $r_{b \, 0}$ the Bohr radius for $\psi = 0 $.
This equation shows that the VSL theory that is being studied is
equivalent, with respect to the hydrostatic equilibrium of Mercury, to
a theory in which the only ``constant" that varies is $G$.  

Now we are in a position to use the results of~\cite{Mc78}. The
variation in the radius of Mercury ($R$) produced by the variation in
$G$ can be parametrized as~\footnote{We call ``$\delta$'' the
  parameter called ``$\alpha$'' in \cite{Mc78} to avoid confusion with the
  fine structure constant.}
\begin{gather}
  \frac{1}{R} \: \frac{dR}{dx_0} = - \: \frac{\delta}{\bar{G}(t)} \:
  \frac{d \bar{G}(t)}{dx_0},\\ 
  \intertext{or equivalently}
  \label{eq:parametrizacion cambio R}
  \frac{1}{R} \: \frac{dR}{dt} = - \: \frac{\delta}{\bar{G}(t)} \: \frac{d
    \bar{G}(t)}{dt} \, ,
\end{gather}
where $\delta$ is generally a function of $ \bar{G} $ and $ M $. Using models
of Mercury, McElhinny et al. obtained the value $\delta = 0.02 \pm
0.005$ for that planet.

\subsection{Specification of the parameters}

We will express the parameters $ b_1, q_1, q_2, q_3 $, and $ q_4 $ that
have been introduced in this work in terms of the parameters $ q $ and
$ b $ of the VSL theory. In our approximation $M=N M_N$ where N is the
total number of nucleons. Equation \eqref{eq:conservacion n} requires
that $N\propto \exp(-q_2\psi)$ while Eq.~\eqref{eq:conservacion masa} implies that
$M\propto\exp (-b\psi)$ and from the discussion in Sec.~2, we see that
$M\propto \exp \left[(q-2-q_2)\psi\right] $ and consequently $q_2=b+q-2$.  From
Eq.~\eqref{eq:variacion densidad} $q_3=q-2+3q=4q-2$ while from
Eq.~\eqref{eq:variacion energia total} $q_4=-q_2+q-2$. Also we have $b_1\equiv
q_1/2 - q_2 + b = -q + 2$ and as we explained before we consider the
$q_1=0$ case.

Finally from Eq.~\eqref{eq:G de t} we conclude that
\begin{equation}
  \bar G (t) = \bar G \: \exp{\left[\left(b_1+q_3+q_4-4+\frac{2}{3}q + 
      \frac{2}{3}\right)\psi \right]}= \bar G \:
  \exp{\left[ \left(\frac{11}{3}q -b - \frac{10}{3}\right)\psi \right]}.
  \label{eq:G de q}
\end{equation}

\subsection{Bound for $\dot{c}/c$}
Replacing the previous expression for $ \bar{G}(t) $ in
Eq.~\eqref{eq:parametrizacion cambio R} one gets
\begin{equation}
  \label{eq:bound1}
  \left( \frac{11}{3} q - b - \frac{10}{3} \right) \dot{\psi}(t) = - \: \frac{1}{\delta} \: \frac{\dot{R}}{R} \simeq 0 \pm 5 \times 10^{-12} \text{y}^{-1} \, ,
\end{equation}
with $ \text{y}^{-1} = 1/ \text{year} $. We have taken $\delta = 0.02 \pm
0.005$ and $\frac{\Delta R}{R} = 0 \pm 0.0004$ \cite{Mc78}, where $ \Delta R $
corresponds to a time interval approximately equal to $ 3.5 \times 10^9 $
years.

This result can be combined with bounds for $ \dot{\alpha}/ \alpha $ that have
been obtained using atomic clocks. We can use e.g. the one obtained in
Ref.~\cite{Sortais00}:
\begin{equation}
  \label{eq:bound2}
  \frac{\dot{\alpha}}{\alpha} = (4.2 \pm 6.9) \times 10^{-15} \text{y}^{-1}.
\end{equation}
In this paper we will consider the $b=0$ case, which gives an upper
bound for all non-negative values of $b$. Moreover, in the VSL theory
$\frac{\dot{\alpha}}{\alpha} = q \dot{\psi}$, then Eq.~$\eqref{eq:bound1}$ can be written as $- \: \frac{10}{3} \dot{\psi} = 0 \pm 5
\times 10^{-12} \text{y}^{-1} - \frac{11}{3} \frac{\dot{\alpha}}{\alpha}$. Comparing this last equation with
Eq.~$\eqref{eq:bound2}$ we see that $- \: \frac{10}{3} \dot{\psi} \approx 0 \pm 5 \times
10^{-12} \text{y}^{-1} $. The conclusive result is that
\begin{equation}
  \frac{\dot{c}}{c} = \dot{\psi} =  0 \pm 2 \times 10^{-12} \text{y}^{-1}.
\end{equation}
This can be rewritten as a bound for the adimensional quantity $\psi' = H_0^{\, -1} \dot{\psi}$ after multiplying by the Hubble time ($H_0^{\, -1}$)~\footnote{We have taken $H_0^{\, -1} \approx 1.5 \times 10^{10}$ years.}:
\begin{equation}
  \psi' =  0 \pm 3 \times 10^{-2}.
\end{equation}

\section{White dwarfs luminosities and the scalar field}

White dwarfs are excellent objects to test any energy injection from a
scalar field \cite{Stothers76,Malin80}. This is due both to
their low luminosity, as well as their extremely high heat
conductivity, making all energy microscopically released to enhance
the total luminosity. Most of them are adequately described by
Newtonian physics and a zero temperature approximation, the latter
hypothesis providing a polytrope type equation of state~(EOS). 

\subsection{The polytropic EOS}
\label{sec:PolyEOS}

Using the subscript $0$ to denote quantities without the presence of
the scalar field, we write the polytrope equation as $p_0=K_0\rho^\gamma$.
Given Eq.~\eqref{eq:presion}, we have
\begin{equation}
  p = p_0(\rho) - b_1\gamma\rho\frac{p_0}{\rho}\psi,
\end{equation}
so we can write
\begin{equation} \label{eq:K}
  p = K_0(1-  b_1\gamma\psi)\rho^\gamma,
\end{equation}
and we recover a polytrope equation of state with a new constant $K_0\to
K = K_0(1-b_1\psi\gamma)$.  The Lane-Emden function with polytrope index
$(\gamma-1)^{-1}$ still applies, and consequently the expressions for the
radius, mass, and internal energy of the star will be the same in terms
of the effective constant $K$ \cite{Chandra57}:
\begin{equation}
\label{eq:polenergy}
  E = -\frac{3\gamma-4}{5\gamma-6}\frac{GM^2}{R} ,
\end{equation}
\begin{equation}
\label{eq:polradius}
  R = f^{1/2}\rho_c^{(\gamma-2)/2}\zeta_1 ,
\end{equation}
\begin{equation} 
\label{eq:Mass}
  M = 4\pi \rho_c^{(3\gamma-4)/2} f^{3/2}
  \zeta_1^2 |\theta'(\zeta_1)| ,
\end{equation}
where $\rho_c$ is the central density, $\zeta_1$ is the first root of the
Lane-Emden function, and we have defined $f=K\gamma/(4\pi G(\gamma-1))$. Unlike with the topographic analysis of planetary palaeoradii, the energy balance in the star is not invariant under the scaling~\eqref{eq:scaling}. Thus, from Eq.~\eqref{eq:Mercurio1} we see that the $G$ for the internal energy of the star is $G(t)=\bar G\exp[(b_1+q_3+q_4-4)\psi]=\bar G\exp[(3q - b -4)\psi]$. On the other hand, the dependence of the radius $R$ on $\psi$ can be obtained solving Eq.~\eqref{eq:Mass} for the central density and replacing it in Eq.~\eqref{eq:polradius} [the $\psi$ dependences of the total mass and the compressibility were given in Sec.~6.2 and in Eq.~\eqref{eq:K}, respectively]. Finally Eq.~\eqref{eq:polenergy} leads to the following dependence of the internal energy:
\begin{equation}
  E \propto \exp{\psi f(q,b,\gamma)} ,
\end{equation}
where
\begin{equation}
 f(q,b,\gamma)\equiv 3 q - b - 4 + \frac{2\gamma-4 + b(5-5\gamma)+q(
3-\gamma)}{3\gamma-4}.
\end{equation}

To go on we assume that the star is in equilibrium in the sense that all the energy injected by the field $\psi$ is radiated away. Therefore the equation for the luminosity induced by the $\psi$ field $(L_\psi)$ becomes
\begin{equation}
\label{eq:lpsi}
  L_\psi= -\dot E = -f(\gamma,q,b) E \dot\psi.
\end{equation}

\subsection{Comparison with observation}
\label{sec:WDObs}

We shall consider only white dwarfs well described by the nonrelativistic value $\gamma=5/3$, in which case $f(q,b,\gamma)=\frac{13}{3} q-
\frac{13}{3} b -\frac{14}{3}$. Unfortunately, only 
a handful of white dwarfs have well-measured masses, radii, and
luminosities: indeed, these four or five stars are used to test the
theory of white dwarfs, since the mass-radius relation requires
exactly the same parameters we need to carry our comparison of theory
and experiment.
These stars and their properties have been reviewed in Ref.~\cite{provencal02}. Table~1 shows the adopted values. We have
excluded Sirius B from the sample, since relativistic effects are
important in this case.

To obtain the bound for $\dot \psi$ we rewrite Eq.~\eqref{eq:lpsi} as
\begin{equation}
\dot \psi = \frac{L_\psi}{E} \frac{3}{13b+14}+ \frac{13}{13b+14}\frac{\dot\alpha}{\alpha} \; .
\end{equation}  
Using again the upper bound \eqref{eq:bound2} for the present time
variation of $\alpha$ and bounding $L_\psi$ by the observed luminosity of the white dwarfs $(L)$ we obtain 
\begin{eqnarray}
\label{eq:bound3}
  \abs{\dot \psi} &\leq& \frac{L}{E} \; \absb{\frac{3}{13b+14}} + 
\absb{\frac{13}{13b+14}\frac{\dot\alpha}{\alpha}}
  \\ \nonumber
  & \leq & \dot\psi_0 \; \absb{\frac{3}{13b+14}} + \absb{\frac{13}{13b+14}} \; 1.1\times 10^{-14}{y^{-1}},
\end{eqnarray}

where
\begin{equation}
  \dot\psi_0\equiv \frac{(L/L_\odot)}{(E/E_0)} \frac{1}{\tau_\odot},
\end{equation}
and
\begin{equation}
  E_0 = -\frac{3}{7}\frac{GM_\odot^2}{R_\odot},
\end{equation}
is the would be internal energy of the Sun were it described by a
Newtonian $\gamma=5/3$ polytrope. $L_\odot$ is the solar luminosity and
\begin{equation}
  \tau_\odot = \frac{E_0}{L_\odot}\simeq 1.32\times 10^7 y,
\end{equation}
is the Kelvin-Helmholtz solar contraction time scale.  

The table shows upper bounds for $\dot\psi$ in
$y^{-1}$ again for $b=0$.
\begin{table}
  \centering
  \begin{tabular}{|l|c|c|c|c|c|c|} \hline Object & $M/M_\odot$ & $R/R_\odot$ &
  $L/L_\odot$ & $E/E_0$ & $\dot\psi_0$ & $|\dot\psi|\leq$ \\ \hline
  Procyon B & $0.602$ & $0.0123$ &$ 5.8\times 10^{-4}$ & $29.5$ & $1.5\times 10^{-12}$ &$3.3\times 10^{-13}$ \\
  40 Eri B& $0.501$ & $0.0136$ &$ 0.014$ & $18.5$ & $5.7\times 10^{-11}$ &$1.2\times 10^{-11}$ \\
  Stein 2015B& $0.66$ & $0.011$ &$ 3.1\times 10^{-4}$ & $39.6$ & $5.9\times 10^{-13}$ &$1.4\times 10^{-13}$ \\
  \hline
\end{tabular}

  \caption{Data and bounds for selected white dwarfs (data from
    Ref. \cite{provencal02}). } 
  \label{tab:WhiteD}
\end{table}

Stein 2015B provides the strongest bound. Using a value for the Hubble
time $H_0^{-1}\propto 1.5\times 10^{10} y$ we obtain
\begin{equation}
  \frac{1}{H_0}\frac{\dot c}{c} = \frac{1}{H_0}\dot\psi = 0\pm  2.1\times 10^{-3}.
\end{equation}

Comparing the expressions~\eqref{eq:bound1} and \eqref{eq:bound3} we see that white dwarf physics provides the strongest constraints on
the VSL theory near the present epoch for almost all values of the $b$ parameter, except for those near $b=-14/13$ which make Eq.~\eqref{eq:bound3} uninformative. It is also clear that combining both bounds~\eqref{eq:bound1} and \eqref{eq:bound3} we can obtain a bound for $|\dot \psi|$ independent from the value of $b$ (although less strong than the bound given for $b \ge 0$): $|\dot \psi| \leq 2.2 \times 10^{-12} \text{y}^{-1}$.
 
\section{Conclusions}
We have obtained the equations that describe a perfect fluid in the
nonrelativistic limit and the first law of thermodynamics in the
context of the covariant VSL theory proposed by J. Magueijo. We showed that the field
$\psi$ can formally be considered as a new thermodynamic variable and we also showed how to obtain the equations
of state in the VSL theory  when the corresponding equations for constant $c$ are given.

The nonrelativistic hydrostatic equilibrium equation has the usual
form with the gravitational constant $G$ replaced by an effective
constant. The different variables (pressure, mass, mass density)
depend on the $\psi$ field and so the radius of a planet should vary in
time. Using bounds for the variation of the
radius of Mercury and the fine structure constant we have set limits
on $ \dot{c}/c: \; \dot{c}/c=\dot{\psi} = 0 \pm 2 \times 10^{-12}
\text{y}^{-1}$ (valid for positive values of the $b$ parameter). The most interesting thing about this result is that
it gives a bound for $\dot{\psi} ,$ whereas the known limits for the
variation of $\alpha$ and $e$ lead to bounds for the product $ q \dot{\psi}$. 

Under the same Newtonian approximation we obtained the dependence of
the luminosity of a white dwarf on the time variation of the scalar
field. The bound obtained is more stringent than the planetary radius
bound by an order of magnitude. 

The $b=0$ assumption suggests a null coupling of the scalar field with
matter. However, the $q\neq 0$ assumption implies a quantum coupling
between $\psi$ and matter, not explicitly shown in the action
\eqref{eq:accionMag}. Of course this and other issues such as the
microscopic origin of the energy exchange between the scalar field and
ordinary matter as well as whether all the energy injected by $\psi$ on a
star is radiated away or not, deserve further work. This we leave for
future communications.
\section{Appendix: spatial and temporal behavior of $\psi$}
After some simplifications valid in the Newtonian limit, the equation
for $\psi$ (in the case $a \neq 0$) becomes:
\begin{equation}
  \Box \psi  = \frac{8 \pi G}{c^4 (2\kappa + 3a^2)} \; aT  \, .
\end{equation}
Since $\psi$ is small, $c^4$ can be replaced by $c_0^{\; 4}$ in the right-hand side of the equation (this is the first step of an iterative
process of resolution):
\begin{equation}
  \label{eq:simplificacion psi 2}
  \Box \psi  = \frac{8 \pi G}{c_0^{\; 4} (2\kappa + 3a^2)} \; aT  \, .
\end{equation}
$ T $ can be written as the sum of a spatial part $ T_s $
(corresponding to Mercury) and a temporal one $ T_t $ (cosmological
term). Then $\psi $ can also be separated into spatial ($ \psi_s $) and
temporal ($ \psi_t $) components. The metrics to be used are the quasi-Minkowskian and the cosmological (Robertson-Walker) for the spatial
and temporal parts, respectively.
\subsection{Spatial behavior}
Using that $ T \simeq \rho = $ energy density of Mercury, the spatial part of
Eq.~$\eqref{eq:simplificacion psi 2}$ can be written:
\begin{equation}
  \nabla^2\psi_s = \frac{1}{r^2} \: \frac{d}{dr} \left( r^2 \frac{d \psi_s}{dr} \right) =  \frac{8 \pi G}{c_0^{\; 4}} \: \frac{a}{2\kappa + 3a^2} \rho \, .
\end{equation}
After integrating one arrives at
\begin{equation}
  \frac{d \psi_s}{dr} = \frac{2 G}{c_0^{\; 4}} \: \frac{a}{2\kappa + 3a^2} \frac{U(r)}{r^2} \, .
\end{equation}
This is one of the formulas that has been used in our work. It can
help us to obtain an estimation of the variation of $\psi$ inside
Mercury:
\begin{equation}
  \Delta \psi \approx \frac{2a}{2\kappa + 3a^2} \: \frac{4}{3} \pi \frac{1}{2} \left( \frac{G}{c_0^{\; 2}} \bar{\rho_m} R^2 \right),
\end{equation}
where $\bar{\rho_m}$ and $R$ are the average density and the radius of
Mercury, respectively. The first factors are presumably $O(1)$ so
\begin{equation}
  \Delta \psi \approx \frac{G}{c_0^{\; 2}} \bar{\rho_m} R^2 \approx 10^{-11} \, .
\end{equation}
This number is small compared to the bound that is obtained for the
variation of $\psi$ in the last $ 3.9 \times 10^9 $ years (the relevant period
of time for our study of Mercury) and besides the temporal and spatial
behavior of this field can be separated. This justifies the steps of
our analysis in which $\psi$ was considered constant inside Mercury.
\subsection{Temporal behavior}
The temporal part of Eq.~$\eqref{eq:simplificacion psi 2}$ is
\begin{equation}
  \Box \psi_t = \frac{8 \pi G}{c_0^{\; 2}} \: \frac{a}{2\kappa + 3a^2} \rho_{mU} \, ,
\end{equation}
where $ \rho_{mU} $ is the average density of the Universe. Using the
Robertson-Walker metric it becomes
\begin{equation}
  \label{eq:psi temporal}
  \nabla_0 \nabla^0 \psi = \frac{8 \pi G}{c_0^{\; 2}} \: \frac{a}{2\kappa + 3a^2} \rho_{mU} +
  3 \frac{\dot{a}(t)}{a(t)} \dot{\psi}, 
\end{equation}
where $a(t)$ is the scale factor of the metric and $a$ is one of the
parameters of the VSL theory. In the way towards obtaining the
gravitational potential equation in the Newtonian limit, the equation
$G_{0}^{\; 0} = R_{0}^{\; 0} - \frac{1}{2} R = \frac{8 \pi G}{c^4}\;
T_{0}^{\; 0} + a \left( \nabla_0 \nabla^0 \psi - \Box \psi \right)$ appears. We want to
show that $ \nabla_0 \nabla^0 \psi $ is negligible compared to the order $\psi$ term
in the Taylor's expansion of $\frac{8 \pi G}{c^4}\; T_{0}^{\; 0}$. To do
this it will be demonstrated that each term appearing in
Eq.~$\eqref{eq:psi temporal}$ can be neglected:
\begin{itemize}
\item First term of Eq.~$\eqref{eq:psi temporal}$:
  This term is negligible since $T_{0}^{\; 0} \approx \rho_{\text{Mercury}}
  c^2$ and $\frac{\rho_{mU}}{\rho_{\text{Mercury}}} \approx 10^{-30}$.
\item Second term of Eq.~$\eqref{eq:psi temporal}$:
  \begin{equation}
    3 \frac{\dot{a}(t)}{a(t)} \dot{\psi}  \approx 3 \frac{H_0}{c_0^{\; 2}} \:
    \frac{\Delta \psi}{\Delta t} = 3 \frac{1}{c_0^{\; 2} \, H_0^{\, -1} \, \Delta t}
    \Delta{\psi} \, , 
  \end{equation}
  where $\Delta \psi $ represents the change of $\psi$ in a time interval $\Delta t$. Taking
  $\Delta t \approx 4 \times 10^9 $ years (this is the time interval for which we have
  a bound for the variation of the radius of Mercury), we get
  \begin{equation}
    3 \frac{\dot{a}(t)}{a(t)} \dot{\psi} \approx 10^{-46} \,\Delta \psi \; \text{cm}^{-2}.
  \end{equation}
  This quantity must be compared with
  \begin{equation}
    (-4) \: \frac{8 \pi G}{c_{0}^{\; 4}} \; \rho \psi \approx 10^2 \times
    \frac{G}{c_{0}^{\; 2}} \times \rho_m \times \psi \approx 10^{-26} \psi \; \text{cm}^{-2} \, ,
  \end{equation}
  and hence we see that the second term of Eq.~$\eqref{eq:psi temporal}$ is also
  negligible.
\end{itemize}

\end{document}